\documentclass[10pt,conference]{IEEEtran}

\usepackage{hyperref}

\usepackage{comment}
\usepackage{booktabs}
\usepackage{lineno}
\usepackage{makecell}
\usepackage{amsmath,amsfonts}
\usepackage{algorithmic}
\usepackage{graphicx}
\usepackage{textcomp}
\def\BibTeX{{\rm B\kern-.05em{\sc i\kern-.025em b}\kern-.08em
    T\kern-.1667em\lower.7ex\hbox{E}\kern-.125emX}}

\usepackage{bm}
\usepackage{listings}
\usepackage[T1]{fontenc}    
\usepackage{xcolor}
\usepackage[dvipsnames]{xcolor}
\usepackage[most]{tcolorbox}

\usepackage{soul}
\usepackage{makecell}

\usepackage{pifont}

\newcommand{\tool}[0]{\textsc{JsonATG}}

\usepackage[shortlabels]{enumitem}
\setlist[itemize]{align=parleft,left=0pt..1em}
\setlist[enumerate]{align=parleft,left=0pt..1.5em}

\usepackage{amsthm}
\newtheorem{definition}{Definition}

\definecolor{pythonblue}{rgb}{0.16,0.12,0.93}
\definecolor{cppgreen}{rgb}{0.16,0.42,0.16}
\definecolor{promptinsert}{HTML}{bfefff}
\definecolor{compcolor}{HTML}{90EE90}
\definecolor{codehlcolor}{HTML}{ffec8b}
\definecolor{codehlcolor2}{HTML}{ffbbff}
\definecolor{bgcolor}{rgb}{0.95,0.95,0.92}
\definecolor{backcolour}{rgb}{0.95,0.95,0.92}
\definecolor{codegreen}{rgb}{0,0.6,0}
\definecolor{commentgreen}{RGB}{2,112,10}
\definecolor{codegray}{rgb}{0.5,0.5,0.5}
\definecolor{codepurple}{rgb}{0.58,0,0.82}

\newcommand{\myjavafont}{\fontfamily{SourceCodePro-TLF}\footnotesize}

\lstdefinestyle{java}{
    language=Java,
    basicstyle=\myjavafont,
    keywordstyle=\color{codepurple},
    emph={JSON,JSONObject,JSONArray,JSONPath},
    emphstyle={\color{blue}},
    numberstyle=\tiny\color{codegray},
    stringstyle=\color{commentgreen},
    numbers=left,
    numberstyle=\tiny\color{black},
    stepnumber=1,
    numbersep=7pt,
    backgroundcolor=\color{cyan!8},
    tabsize=2,
    captionpos=b,
    breaklines=true,
    breakatwhitespace=false,
    showspaces=false,
    showstringspaces=false,
    showtabs=false,
    frame=single,
    rulecolor=\color{black},
    xleftmargin=2ex,
    xrightmargin=1em,
    framesep=4pt,
    belowskip=4pt
}

\lstdefinestyle{python}{
    language=Python,
    basicstyle=\fontsize{8}{10}\ttfamily,
    keywordstyle=\color{blue},
    commentstyle=\color{gray},
    stringstyle=\color{black},
    showstringspaces=false,
    breaklines=true,
    breakindent=0pt,
    breakatwhitespace=false,
    escapeinside={(*@}{@*)}
}

\lstdefinestyle{cpp}{
    language=C++,
    basicstyle=\fontsize{8}{10}\ttfamily,
    keywordstyle=\color{blue},
    commentstyle=\color{gray},
    stringstyle=\color{green},
    showstringspaces=false,
    breaklines=true,
    breakindent=0pt,
    breakatwhitespace=false,
    escapeinside={(*@}{@*)}
}

\lstdefinestyle{plain}{
    basicstyle=\fontsize{8}{10}\ttfamily,
    keywordstyle=\color{blue},
    commentstyle=\color{gray},
    stringstyle=\color{green},
    showstringspaces=false,
    breaklines=true,
    breakatwhitespace=false,
    breakindent=0pt,
    escapeinside={(*@}{@*)}
}

\lstdefinestyle{python2}{
    language=Python,
    basicstyle=\fontsize{8}{10}\ttfamily,
    keywordstyle=\color{blue},
    commentstyle=\color{gray},
    stringstyle=\color{green},
    showstringspaces=false,
    breakatwhitespace=false,
    breaklines=true,
    breakindent=0pt,
    escapeinside={(*@}{@*)}
}

\lstdefinestyle{cpp2}{
    language=C++,
    basicstyle=\fontsize{8}{10}\ttfamily,
    keywordstyle=\color{blue},
    commentstyle=\color{gray},
    stringstyle=\color{green},
    showstringspaces=false,
    breaklines=true,
    breakindent=0pt,
    breakatwhitespace=false,
    escapeinside={(*@}{@*)}
}

\lstdefinestyle{sql}{
    language=SQL,
    basicstyle=\fontsize{8}{10}\ttfamily,
    keywordstyle=\color{blue},
    commentstyle=\color{green},
    stringstyle=\color{black},
    showstringspaces=false,
    breakatwhitespace=false,
    breaklines=true,
    breakindent=0pt,
    escapeinside={(*@}{@*)}
}

\lstdefinestyle{prompt}{
    language=Python,
    basicstyle=\fontsize{8}{10}\ttfamily,
    keywordstyle=\color{blue},
    commentstyle=\color{gray},
    showstringspaces=false,
    breaklines=true,
    keepspaces=true, 
    breakindent=0pt,
    breakatwhitespace=false,
    showspaces=false,   
    escapeinside={(*@}{@*)}
}
\lstdefinestyle{text}{
    basicstyle=\myjavafont,
    showstringspaces=false,
    breaklines=true,
    breakatwhitespace=false,
    breakindent=0pt,
    keepspaces=true,
    showspaces=false,   
    escapeinside={(*@}{@*)},
    xleftmargin=1em,
    xrightmargin=1em,
    framexleftmargin=0pt, % 调整代码与左侧边框线之间的距离
    framexrightmargin=0pt, % 调整代码与右侧边框线之间的距离
    framesep=0pt,
    belowskip=3pt
}

\newcommand{\inserthl}[1]{\sethlcolor{promptinsert}\hl{#1}}
\newcommand{\comphl}[1]{\sethlcolor{compcolor}\hl{#1}}
\newcommand{\codehl}[1]{\sethlcolor{codehlcolor}\hl{#1}}
\newcommand{\codehlerr}[1]{\sethlcolor{codehlcolor2}\hl{#1}}

\makeatletter % changes the catcode of @ to 11
\newcommand{\linebreakand}{%
  \end{@IEEEauthorhalign}
  \hfill\mbox{}\par
  \mbox{}\hfill\begin{@IEEEauthorhalign}
}
\makeatother % changes the catcode of @ back to 12

\begin{document}

\title{Multi-Agent Assisted Automatic Test Generation for Java JSON Libraries}

\IEEEspecialpapernotice{\vspace{-4.2ex}}
\author{
\begin{tabular}{cc}
  \begin{minipage}{0.48\textwidth}
    \centering
    Sinan Wang$^*$\\
    \normalsize\textit{Research Institute of Trustworthy Autonomous Systems}\\
    \textit{Southern University of Science and Technology}\\
    Shenzhen, Guangdong, China\\
    wangsn@mail.sustech.edu.cn
  \end{minipage}
&
  \begin{minipage}{0.48\textwidth}
    \centering
    Zhiyuan Zhong$^{*,1}$\\
    \normalsize\textit{School of Computing}\\
    \textit{University of Utah}\\
    Salt Lake City, UT, USA\\
    zhiyuan.zhong@utah.edu
  \end{minipage}
\\[8ex]
  \begin{minipage}{0.48\textwidth}
    \centering
    Shaojin Wen\\
    \normalsize\textit{Alibaba Group}\\
    Hangzhou, Zhejiang, China\\
    shaojin.wensj@alibaba-inc.com
  \end{minipage}
&
  \begin{minipage}{0.48\textwidth}
    \centering
    Yepang Liu$^2$\\
    \normalsize\textit{Department of Computer Science and Engineering}\\
    \textit{Southern University of Science and Technology}\\
    Shenzhen, Guangdong, China\\
    liuyp1@sustech.edu.cn
  \end{minipage}
\end{tabular}
}
\maketitle

\begingroup\renewcommand\thefootnote{*}
\footnotetext{The first two authors contributed equally to this work.}
\endgroup

\begingroup\renewcommand\thefootnote{1}
\footnotetext{Work conducted during Zhiyuan Zhong's undergraduate studies at Southern University of Science and Technology.}
\endgroup

\begingroup\renewcommand\thefootnote{2}
\footnotetext{Yepang Liu is the corresponding author. He is also affiliated with the Research Institute of Trustworthy Autonomous Systems.}
\endgroup

\begin{abstract}

JSON is a widely used data format for data exchange between application systems and programming frontends.
In the Java ecosystem, \textit{Java JSON libraries} serve as fundamental toolkits for processing JSON data, powering a wide range of real-world applications, such as web services, Android apps, or data management systems.
However, without effective quality assurance methods, such as automatic test generation (ATG), developers risk introducing subtle data inconsistency bugs, compatibility issues, and even security vulnerabilities.
These flaws can affect billions of end users and potentially cause severe financial losses.

Recently, large language models (LLMs) have shown strong potential in enhancing ATG.
However, existing LLM-based methods, such as \textsc{TitanFuzz} or \textsc{YanHui}, lack specialization in the JSON domain.
For Java JSON libraries, effective bug-triggering test cases should capture the constraints between formatted data and application programs, posing challenges to existing methods, leaving critical aspects of quality assurance unaddressed.

To fill this gap, we propose \tool, a multi-agent ATG system that generates diverse bug-triggering tests for Java JSON libraries.
With historical bug-triggering unit tests as seeds, {\tool} introduces a code summarization agent and a test validation agent into the generation pipeline to produce new, valid test cases.
It applies agent-generated program mutation rules tailored specifically for the structural and semantic characteristics of Java JSON libraries, such as data streaming operations, serialization formats, and data-binding patterns.
The generated tests are further refined through post-processing to ensure both syntactic and semantic correctness.
Our experiments show that {\tool} achieves higher coverage than two state-of-the-art LLM-based test generation methods on the widely used Java JSON library.
Using {\tool} with a \$25 budget, we reported 59 bugs (including non-crashing functional bugs) in \textit{fastjson}, of which 47 were confirmed and 28 have already been fixed.

\end{abstract}


\section{Introduction}

Data-serialization libraries are crucial tools in modern software development.
They are responsible for conversion between programmable in-memory objects and data persistence formats, enabling efficient data exchange, storage, and retrieval for application programs.
Well-known data-serialization formats include XML, JSON, Protocol Buffers, and so on.
In the Java ecosystem, developers widely use JSON in diverse application scenarios, including web services, mobile apps, data management, and scientific computing~\cite{harrand2021behavioral}.
We refer to the utilized programming toolkit as \textit{Java JSON library}, which offers essential functionalities like parsing JSON into Java objects, serializing objects into JSON format, and manipulating JSON data.
Despite their widespread use, Java JSON libraries can suffer from serious bugs, including performance downgrading~\cite{fastjsonCVE} and even security threats~\cite{orgjsonCVE}, potentially affecting the reliability and safety of the dependent applications.

The research community has actively explored automated approaches to ensure the quality of such third-party libraries.
Fuzzing~\cite{chen2023hopper} is a powerful technique that uncovers bugs by generating diverse inputs to exercise software behaviors.
However, traditional fuzzing methods often fall short in detecting non-crashing functional bugs that lead to incorrect outputs without triggering unhandled exceptions.
Moreover, the generated inputs are typically not reproducible or maintainable, which limits their long-term utility in regression testing.
Unit testing allows developers to provide human-written test cases to ensure the correct implementation of the software under test.
Researchers have also proposed to automatically generate such tests~\cite{sun2022mining,almasi2017industrial}, which we refer to as \textit{automatic test generation} (ATG).
However, these techniques cannot be directly applied to Java JSON libraries.
A key limitation is their lack of domain knowledge, such as evaluating the correctness of various serialization configurations or understanding the mappings between JSON data and its corresponding schema (i.e., the Java bean class).
As a result, their ability to expose subtle misbehaviors in Java JSON libraries is greatly constrained.

Recent advances in large language model (LLM) technology have shown promise in ATG~\cite{yuan2024evaluating,schafer2023empirical,chen2024chatunitest,siddiq2024using,deljouyi2025leveraging}.
For example, \textsc{TitanFuzz}~\cite{deng2023large} synthesizes test programs for deep learning libraries with two types of LLM: a generative LLM for creating seed programs and an infilling LLM for rewriting existing programs.
\textsc{YanHui}~\cite{guan2024large} is a directed ATG approach toward the model optimization module in deep learning libraries.
% It shows that LLM can understand the model optimization bugs from historical artifacts and generalize to detect new ones.
Yet, little work has utilized LLM for addressing the aforementioned challenge in testing Java JSON libraries, leaving the potential of LLM in this domain unexplored.

To bridge this gap, we present \tool, a multi-agent ATG system tailored to Java JSON libraries.
{\tool} leverages historical bug-triggering test cases and incorporates LLM-based agents for code summarization and test validation.
It generates diverse and semantically rich test cases using agent-generated program mutation rules that capture the structural and behavioral characteristics of JSON processing and data-binding in Java.
Our experiments show that {\tool} achieves the state-of-the-art performance in generating bug-triggering test suites for popular Java JSON libraries.
In summary, this work makes the following contributions:
\begin{itemize}[itemsep=1pt,align=parleft]
\item We decompose the overall ATG process for Java JSON libraries into three well-defined subtasks and assign each to a dedicated agent, which emulates the workflow of expert human testers in a specialized domain.
\item We implement our proposed multi-agent approach as an ATG system, \tool, and experimentally show that it outperforms two state-of-the-art LLM-based ATG methods in terms of code coverage and detected bugs.
\item With \tool, we have reported 59 bugs (including non-crashing functional bugs) in \textit{fastjson}, a popular Java JSON library, with a gross budget of \$25. Among the reported bugs, 47 were confirmed and 28 have already been fixed.
\end{itemize}

\section{Background}

\subsection{Large Language Models for Code}

Recent developments have demonstrated the effectiveness of LLMs in code-related tasks, including program synthesis~\cite{austin2021program}, repair~\cite{xia2023automated}, debugging~\cite{chen2024teaching}, fuzzing~\cite{xia2024fuzz4all}, test improvement~\cite{alshahwan2024automated}, and decompilation~\cite{xie2024resym}.
Their performance is enhanced with prompt engineering and fine-tuning techniques.
Few-shot prompting~\cite{brown2020language}, using curated examples, guides LLMs in solving tasks with desired output.
Chain-of-Thought (CoT) prompting~\cite{wei2022chain} helps to break down complex problems into smaller steps. 
Embedding relevant code snippets, documentation, and instructions leverages models' in-context learning ability for diverse coding scenarios.
Fine-tuning adjusts the model weights to optimize performance in specific tasks, but is known to be a resource-intensive approach~\cite{wei2024magicoder}.
The LLM-based multi-agent system (LMA) extends the capabilities of single-LLM approaches by integrating multiple collaborating LLMs (referred to as \textit{agents}) to solve complex code generation tasks \cite{dong2024self}.
This is achieved by decomposing the overall task into role-specific sub-goals and delegating each sub-goal to a specialized agent, enabling coordinated problem-solving through division of labor (DoL) \cite{he2025llm}.

\subsection{Automatic Test Generation}

Automatic test generation (ATG) techniques aim to automatically create tests for a given software library.
The output of an ATG approach is typically a \textit{test suite} that consists of multiple \textit{test cases}.
Each generated test case has a \textit{prefix} that sets the conditions necessary to execute the library API method, and a \textit{oracle} (i.e., assertions) that describes the method's expected behavior~\cite{yuan2024evaluating}.
By exploring diverse combinations of execution paths, running configurations, and test inputs, the ATG approaches comprehensively exercise the software under test to uncover potential bugs, verify functional correctness, and improve code coverage.

Traditional ATG approaches often employ random algorithms~\cite{pacheco2007randoop} or search-based strategies~\cite{fraser2011evosuite} to synthesize test cases, or adopt model-based methods to accomplish comprehensive functional coverage~\cite{enoiu2016automated}.
More recent research has explored the use of deep learning (DL) to improve ATG \cite{dinella2022toga,watson2020learning}.
The emergence of LLM has further advanced this area, enabling ATG through prompting to code language models~\cite{yuan2024evaluating,chen2024chatunitest,schafer2023empirical}.

\subsection{Java JSON Library}

JSON (JavaScript Object Notation) is a lightweight, text-based data interchange format.
It was formally standardized as ECMA-404 \cite{ecmainternationalECMA404Ecma} and later by RFC 8259 \cite{ietf8259JavaScript}, which defines its syntax and semantic rules, such as allowed data types, string encoding, and number formatting.

\begin{figure}[t]
  \centering
  \includegraphics[width=0.99\columnwidth]{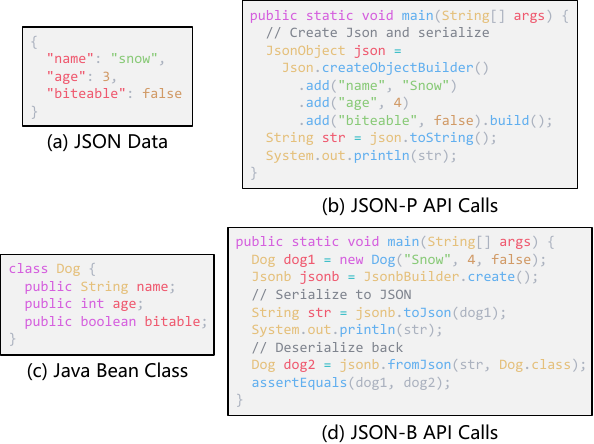}
  \caption{Examples of JSON-P and JSON-B APIs}
  \label{fig:jsonbph}
\end{figure}

\begin{figure*}[t]  
  \centering
  \includegraphics[width=0.99\textwidth]{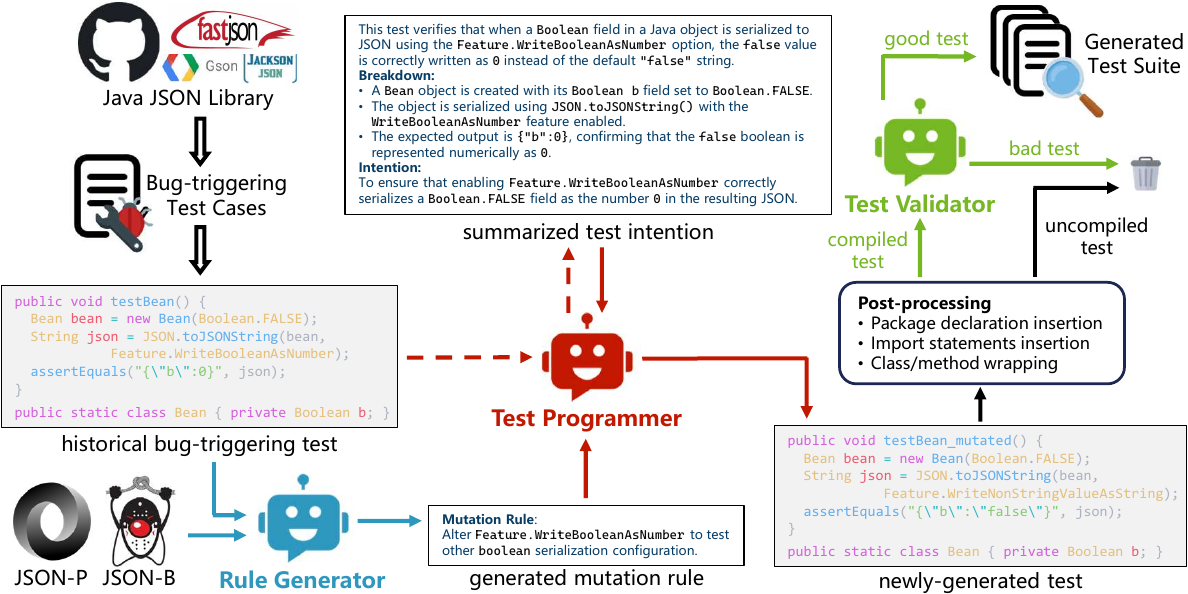}
  \caption{Overview of \tool}
  \label{fig:fuzz_overview}
\end{figure*}

Java developers often use JSON for data exchange across various services and applications.
To facilitate JSON serialization and deserialization, the Java ecosystem has a wide range of JSON libraries.
Among them, \textit{fastjson}~\cite{fastjson2}, \textit{Gson}~\cite{gson}, and \textit{Jackson}~\cite{jackson} are some of the most widely used third-party libraries.
They provide rich functionality for parsing JSON strings into Java objects and vice versa, handling common data types, nested structures, and custom serialization configuration.
However, their implementations (API names, usages, compliance with formal standards, etc.) often diverge in subtle ways.
To bring consistency to JSON handling in Java, two official standards were introduced in the Java community:
\begin{itemize}
\item \textbf{JSON-P} (Java API for JSON Processing) provides low-level streaming and object-model APIs for parsing, generating, and manipulating JSON data. It serves as the foundational reference for JSON parsers in the Java ecosystem.
\item \textbf{JSON-B} (Java API for JSON Binding) offers a high-level, annotation-driven framework for binding native Java Bean \cite{white2005simplifying} objects to and from JSON data.
By standardizing object conversion behaviors, JSON-B enables more convenient and consistent data handling within the Java platform.
\end{itemize}
Fig.~\ref{fig:jsonbph} shows examples of JSON-P and JSON-B API usages.
These standards serve as reference implementations of Java JSON libraries.
Due to the flexibility of third-party libraries, they do not fully conform to the official standards.
However, these standards demonstrate general patterns in using JSON data format with the Java language, which allow us to design a generalizable ATG approach among the Java JSON libraries.

\section{\tool}

Fig.~\ref{fig:fuzz_overview} illustrates the technical overview of our proposed approach, \tool.
{\tool} relies on two external resources: the repository of a Java JSON library and the documentation of the Java JSON API standards, namely JSON-P and JSON-B.
It incorporates three LLM-based agents, each named to reflect its specific role within the overall ATG workflow:
The \textbf{Rule Generator} takes as input a historical bug-triggering test along with the relevant JSON-P or JSON-B API documentation, and produces a mutation rule written in natural language.
The \textbf{Test Programmer} first summarizes the intention of the historical test, then uses this summary with the agent-written mutation rule to generate a new test case. This new test modifies the original test program according to the rule's description.
The newly-generated test is then post-processed to correct any syntactic errors and is passed to the \textbf{Test Validator}, who assesses its logical validity.
The final output of {\tool} is the set of logically valid tests, which we refer to as \textit{good tests} (see Section \ref{ssec:test-validator}).
The remainder of this section details each agent's design.

\subsection{The Mutation Rule Generator Agent}
\label{ssec:rule-generator}

Effective mutation rules help expose domain-specific bugs in the library.
For instance, \textsc{YanHui}~\cite{guan2024large} employs prompt-based mutation rules derived from error-prone APIs, data types, and values.
However, \textsc{YanHui}'s mutation rules are manually summarized, which limits their generalizability and scalability in the new domain.
Instead, given the reference implementation of Java JSON libraries (JSON-P and JSON-B) that describe their common functionalities, we assign the task of generating mutation rules to a dedicated agent.
This enables the automatic generation of diverse and potentially unbounded mutation rules tailored to the JSON domain.

\begin{figure}[t]
\resizebox{\columnwidth}{!}
{\input{prompts/rule_generator.tex}}
\caption{Context of the Mutation Rule Generator agent}
\label{fig:rule-generator}
\end{figure}

The Mutation Rule Generator agent is responsible for writing mutation rules from existing test cases to generate new, semantically meaningful tests.
Fig.~\ref{fig:rule-generator} shows the prompting context for generating the mutation rule.
The process is divided into two consecutive LLM queries.

In the first LLM query, we provide the agent with the historical bug-triggering test case to be mutated, along with a list of API classes from the JSON-P and JSON-B standards.
Recall that JSON-P and JSON-B are two official Java API standards that exemplify typical usage patterns for JSON processing in Java, capturing common functionalities that are broadly applicable across different Java JSON libraries.
Specifically, JSON-P includes 31 classes, while JSON-B has 26 classes.
The agent is instructed to identify the most relevant classes from these standard classes that are associated with the given test case.
For the motivating example test illustrated in Fig.~\ref{fig:fuzz_overview}, the LLM identifies \texttt{JsonWriter} from JSON-P and \texttt{JsonbSerializer} from JSON-B as the most relevant classes, as both are related to the serialization functionality commonly used in JSON processing.

In the second query, we extract the documentation of the identified relevant classes and prompt the agent to generate a possible mutation rule based on the provided descriptions.
To further constrain the output format, we include several example mutation rules as in-context demonstrations \cite{brown2020language} for the LLM.
These examples include:
\begin{itemize}
\item \textbf{Extra Data Retrieval}: add additional getter calls and assert the returned value to validate data integrity.
\item \textbf{Alter Deserialization}: change deserialization method to cover overloaded APIs, such as replacing the string input with a byte array argument.
\item \textbf{Configuration Option}: switch the method argument to test different serialization or deserialization configurations.
\item \textbf{Bean Modification}: change field names and types, add or remove fields, or modify nested classes in the bean class.
\end{itemize}

While syntax-level mutation \cite{ou2024mutators} is feasible, it is hard to adjust the corresponding assertions.
LLM can capture dependencies between program elements and thus implicitly handle the adjustment of test oracle~\cite{hossain2025togll}.
Therefore, instead of applying mutation rules at the syntax level, we will ask the LLM (i.e., the Test Programmer agent) to directly perform program mutation on the original test case.

\subsection{The Test Programmer Agent}
\label{ssec:test-programmer}

\begin{figure}[t]
\resizebox{\columnwidth}{!}
{\input{prompts/test_programmer.tex}}
\caption{Context of the Test Programmer agent}
\label{fig:test_programmer}
\end{figure}

Using the same historical bug-triggering test as the Rule Generator, the Test Programmer is responsible for generating a new test case.
Its task comprises two phases:
1) summarizing the intention of the original test (illustrated by the dashed line in Fig.~\ref{fig:fuzz_overview});
and 2) generating a new test that preserves the original intention but introduces slight modifications based on the provided mutation rule (illustrated by the solid line in Fig.~\ref{fig:fuzz_overview}).
The context of the Test Programmer is shown in Fig.~\ref{fig:test_programmer}.

The Test Programmer first takes the original test as input and is prompted to summarize its underlying intention, such as identifying the targeted API method and describing the key operations taken.
This summarized intention serves as in-context information to guide the Test Programmer in understanding the original test, as suggested by~\cite{agarwal2024faithfulness} and~\cite{chen2024teaching}.
An illustrative example is shown in Fig.~\ref{fig:fuzz_overview}, where the original test program is decomposed into smaller logical steps.
The agent also identifies that the test focuses on the serialization configuration option \texttt{Feature.WriteBooleanAsNumber}.
Then, with the generated mutation rule that suggests altering this configuration, the Test Programmer replaces the original configuration with \texttt{Feature.WriteNonStringValueAsString}.
Crucially, with the reasoning capabilities of LLM, the assertion statement is also adapted to reflect the new expected behavior.
As shown in the example, the serialized value associated with the key \texttt{"b"} changes from the number \texttt{0} to the string \texttt{"false"}, which correctly corresponds to the altered setting.

In practice, we observed that the agent-generated test code often contains minor errors that prevent successful compilation.
To mitigate this issue, we introduce a post-processing phase comprising three key fixing strategies:
\begin{enumerate}
\item \textbf{Package declaration insertion}: Many generated tests omit the Java package declaration. To fix this, we automatically insert an appropriate package statement based on the project structure of the target Java JSON library.

\item \textbf{Import Statement Insertion}: Missing import statements—especially for commonly used classes like assertion utilities—can cause compilation failures. We address this by performing static analysis and inserting necessary import statements to fix all unresolved symbols.

\item \textbf{Class and Method Wrapping}: Generated code snippets often lack a surrounding class or method definition. To address this, we wrap each snippet in a properly declared test class and method, ensuring it adheres to the Java syntax and testing framework (e.g., JUnit) conventions.
\end{enumerate}

\subsection{The Test Validator Agent}
\label{ssec:test-validator}

A failed LLM-generated test does not necessarily indicate a fault in the program under test.
In many cases, failures may come from logically incorrect test code itself~\cite{chen2024chatunitest}.
While the former is desirable for uncovering unknown bugs, the latter can mislead developers and waste their debugging efforts.
To differentiate between these scenarios, we categorize test results into two mutually exclusive types:

\begin{figure}[t]
    \centering
    \includegraphics[width=0.85\columnwidth]{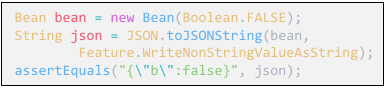}
    \caption{A bad test modified from the motivating example}
    \label{fig:badtest_example}
\end{figure}

\begin{definition}[Good Test]
A good test fails (at an assertion or by throwing an uncaught exception) due to intrinsic bugs in the library. The logic of a good test is considered correct.
\end{definition}
\begin{definition}[Bad Test]
A bad test fails as a result of incorrect test logic or misuse of the library, suggesting that the library works as expected and the test code is faulty.
\end{definition}

The motivating example demonstrated in Fig.~\ref{fig:fuzz_overview} is a good test as it causes a test failure and reveals a real bug in \textit{fastjson} (see Section~\ref{ssec:rq3}).
Fig.~\ref{fig:badtest_example} shows a bad test modified from this good test.
It preserves the test prefix, while the test assertion's expected output is changed: the ``false'' value is changed from a string into a boolean literal.
This modification leads to a logical error because the expected output no longer aligns with the behavior of the \texttt{WriteNonStringValueAsString} feature.
As a result, the test fails not due to a defect in the program under test, but because of an incorrect test oracle.

\begin{figure}[t]
\resizebox{\columnwidth}{!}
{\input{prompts/test_validator.tex}}
\caption{Prompt template of the Test Validator agent}
\label{fig:test_validator}
\end{figure}

For each failed test case, the Test Validator will classify it as either a good test or a bad test.
To improve prediction accuracy, we adopt both few-shot and chain-of-thought prompting techniques~\cite{song2023comprehensive,wei2022chain}.
Following the idea of~\cite{wang2023selfconsistency}, the Test Validator executes three independent LLM queries and determines the final label via majority voting.
The complete prompt template is shown in Fig.~\ref{fig:test_validator}.
Good test examples are collected from the issue tracking systems of the corresponding Java JSON library.
In contrast, bad tests are sampled from LLM-generated failed tests which validated through manual inspection.
In considering the context length limitations of the underlying LLM, each prompt includes two good tests and two bad tests.

\section{Experimental Setup}

\subsection{Subject Library}
\label{ssec:subject}

\textit{Fastjson} is a popular high-performance Java JSON library. %, which outperforms popular alternatives like Jackson and Gson \cite{githubFastjson_benchmark}.
The upgraded version of \textit{fastjson} is \textit{fastjson2}, designed to address performance and security issues while retaining most of the original API names for backward compatibility.
We chose to experiment with \textit{fastjson2} for three reasons:
\begin{enumerate}
\item It is a popular Java JSON library that attracts 4.1K stars on GitHub and 8.4K dependent artifacts on Maven Central;
\item It is actively maintained, with over 2K issues on GitHub, many of which have led to the inclusion of regression unit tests (i.e., classes with naming pattern \texttt{Issue\textbackslash{}d+});
\item The backward compatibility between \textit{fastjson} and \textit{fastjson2} enables the construction of an extra differential oracle for the generated test cases. However, we use this oracle solely to verify detected bugs for ethical considerations~\cite{wang2022aper}.
\end{enumerate}
For simplicity, unless otherwise noted, we refer to \textit{fastjson2} as \textit{fastjson}.
We conducted experiments using \textit{fastjson} version 2.0.49, which was the latest release available at the time.

\subsection{Bug-triggering Test Collection}

Instead of collecting issues or pull requests (PRs) from the \textit{fastjson} repository, we extracted bug-triggering tests directly from its unit test suite.
These tests are typically concise yet contain the essential steps to reproduce specific bugs in the library's APIs.
As they have been reviewed and maintained by the project's contributors, they are generally more reliable.

To identify tests related to historical bugs, we used regular expressions to select Java test files whose filenames reference GitHub issues (e.g., test class named \texttt{Issue100} corresponds to \textit{fastjson}'s~\href{https://github.com/alibaba/fastjson2/issues/100}{issue~100}). This approach yielded 681 historically bug-triggering unit tests, which we used as the experimental dataset for evaluating \tool.

\subsection{Research Questions}

Our experiment aimed to answer the following research questions (RQs) regarding \tool's performance:
\begin{itemize}
\item \textbf{RQ1:} How does {\tool} perform compared to the state-of-the-art LLM-based ATG tools in terms of code coverage?
\item \textbf{RQ2:} How do different components contribute to the overall performance of {\tool}?
\item \textbf{RQ3:} What bugs in \textit{fastjson} can {\tool} discover?
\item \textbf{RQ4:} What are the main causes of bad tests in the generated tests, and how effectively can Test Validator classify them?
\end{itemize}

\subsection{Environments}

Our experiments were conducted on a 64-core workstation running Ubuntu 20.04.6 LTS with two NVIDIA Quadro RTX 6000 GPUs, using OpenJDK 1.8.0-382 and Junit 5.11.

\subsection{LLM Selection for Agents}

We used four representative LLMs for the agents in {\tool} for our experiments.
They included commercial and open-source LLMs:
\begin{itemize}
\item \textbf{DeepSeek} open-sourced a series of coding LLMs that achieve competitive performance among open-source models. We adopted the 6.7b instruction-tuned model. % \textit{deepseek-coder-6.7b-instruct}~\cite{deepseek_hf}.
\item \textbf{Llama3} is a state-of-the-art open-source LLM family. We used the instruction-tuned version with 8b parameters. % \textit{Llama-3-8b-instruct}.
\item \textbf{GPT-3.5} is a proprietary LLM offered by OpenAI, and it is the default model used in the early ChatGPT product. We accessed {GPT-3.5 Turbo} through API services.
\item \textbf{GPT-4o} is a leading commercial model released by OpenAI, recognized for its strong performance on complex and multi-step reasoning tasks.
\end{itemize}

For the Mutation Rule Generator, we adopted DeepSeek as the underlying LLM due to its balance between performance and cost-effectiveness.
For the Test Validator, we employed GPT-4o, selected for its superior reasoning capabilities, which are essential for its binary classification task.
The base model used for the Test Programmer varies according to the RQs under investigation.
In RQ1, we used GPT-3.5 to align the configuration with the two baseline approaches.
In RQ2, we explored the impact of different LLMs on the Test Programmer to assess their contribution to the overall performance of \tool.
In RQ3, we extensively ran the Test Programmer with GPT-3.5 to evaluate the cost of bug detection.

\subsection{Baseline Methods}

For comparison, we chose two state-of-the-art LLM-based ATG approaches as baseline methods:
\begin{itemize}
\item \textbf{\textsc{ChatTester}}~\cite{yuan2024evaluating} is a ChatGPT-based ATG approach.
It consists of a test generator that understands the focal method to produce initial tests and a test refiner that iteratively repairs errors using compiler messages and code context.
\item \textbf{\textsc{ChatUniTest}}~\cite{chen2024chatunitest} is another LLM-based ATG framework that demonstrates reliable coverage across diverse projects. It optimizes the LLM through prompt design, adaptive focal context, and a generation-validation-repair mechanism.
\end{itemize}
There are existing benchmarks for performance testing for Java JSON libraries, such as \textit{JSONTestSuite}~\cite{githubGitHubNstJSONTestSuite} and \textit{Java JSON Benchmark}~\cite{githubGitHubFabienrenaudjavajsonbenchmark}.
However, these benchmarks primarily focus on parsing correctness and throughput, rather than on functional correctness, robustness, or bug-finding capabilities.
Therefore, they are not well-suited for evaluating ATG tools.

\subsection{Evaluation Metrics}

We evaluated {\tool} using two widely adopted metrics: code coverage and the number of newly detected bugs.
Code coverage was measured using JaCoCo~\cite{githubGitHubJacocojacoco} with the instruction-level granularity.
As mentioned in Section~\ref{ssec:subject}, each test was executed on both \textit{fastjson} and \textit{fastjson2} to verify consistency.
Since all seed tests pass on the latest version of \textit{fastjson2}, any newly generated test that fails only on \textit{fastjson2} is considered to reveal a potential bug.
We would review such tests and submit bug reports to \textit{fastjson}'s maintainers to determine whether they uncovered previously unknown bugs.

\section{Results}

\subsection{RQ1: Performance Comparison}
\label{ssec:rq1}

\subsubsection{Method}

We used coverage scores to compare the performance of {\tool} against two state-of-the-art approaches.
To better assess the ability of ATG approaches to exercise critical components of the target library, we collected coverage for six core classes in \textit{fastjson}:
\texttt{JSON}, \texttt{JSONPath}, \texttt{JSONArray}, \texttt{JSONObject}, \texttt{JSONReader}, and \texttt{JSONWriter}.

To run \tool, we employed GPT-3.5 as the underlying LLM for the Test Programmer and instructed it to generate three new test cases for each historical bug-triggering test in our dataset.
For \textsc{ChatTester} and \textsc{ChatUniTest}, we used their open-source implementations from the Maven plugin~\cite{githubGitHubZJUACESISEchatunitestmavenplugin}, with both tools configured to use GPT-3.5 as the underlying LLM.
For each test in our dataset, we extracted the respective focal method and instructed the tools to generate three new test cases.
We allowed up to three rounds of repair, keeping all other parameters at their default values.

\subsubsection{Result}

As shown in Table~\ref{tab:compare_cut_ctr}, \textsc{ChatUniTest} achieved higher coverage for \texttt{JSONObject} and \texttt{JSONArray}, whereas \textsc{ChatTester} had higher coverage for \texttt{JSONObject}.
This can be attributed to the fact that many methods in both \texttt{JSONObject} and \texttt{JSONArray} are simple getter methods. Such methods enable \textsc{ChatUniTest} and \textsc{ChatTester} to easily generate valid test cases that invoke them on initialized instances.
Although these tests contribute to higher coverage, they only retrieve data immediately after object construction and fail to exercise the library's deeper or more complex functionalities.
In contrast, tests generated by {\tool} may not include getters of all data types, resulting in lower coverage scores for these classes.

\begin{table}[t]
\centering
\caption{Comparison with LLM-based approaches \textsc{ChatTester} and \textsc{ChatUniTest}}
\label{tab:compare_cut_ctr}
\begin{tabular}{c c c c}
\toprule
    & \multicolumn{3}{c}{Instruction Coverage (\%)} \\
    \cmidrule(lr){2-4}
Class & \textsc{ChatTester} & \textsc{ChatUniTest} & \tool \\ \midrule
\texttt{JSON}       & $$2.39$$  & $$6.50$$  & \textbf{$$32.76$$} \\
\texttt{JSONPath}   & $$0.33$$  & $$11.16$$ & \textbf{$$20.46$$} \\
\texttt{JSONArray}  & $$27.91$$ & \textbf{$$52.51$$} & $$29.64$$ \\
\texttt{JSONObject} & $$43.94$$ & \textbf{$$47.47$$} & $$41.33$$ \\
\texttt{JSONReader}     & $$1.88$$  & $$4.85$$  & \textbf{$$44.86$$} \\
\texttt{JSONWriter}     & $$2.68$$  & $$4.64$$  & \textbf{$$41.88$$} \\
\bottomrule
\end{tabular}
\end{table}

{\tool} achieved higher coverage scores for four classes: \texttt{JSON}, \texttt{JSONPath}, \texttt{JSONReader}, and \texttt{JSONWriter}.
In contrast, \textsc{ChatUniTest} and \textsc{ChatTester} struggled to cover these larger classes effectively.
This limitation stems from their requirement to include the source code of the focal method in the prompt, which often exceeds the maximum token limit.
Additionally, they face challenges in generating tests for methods with complex usage scenarios, such as the overloaded parse methods with various arguments in \texttt{JSON} and the more intricate methods in \texttt{JSONPath}.
Consequently, they failed to produce valid tests for these classes.

\begin{center}
\noindent\fbox{\parbox{.9\linewidth}{
\textbf{Answer to RQ1.}
{\tool} achieves higher coverage on complex and large classes in the Java JSON library, demonstrating superior capability in exercising critical components of the library under test.
In contrast, state-of-the-art tools struggle with handling complex API usages and prompt length limitations.
}}
\end{center}

\subsection{RQ2: Ablation Study}

\subsubsection{Effectiveness of Mutation Rules}
\label{sssec:mutation-rules}

We first evaluate the effectiveness of applying agent-generated mutation rules.
For comparison, we include a free-mutation variant~\cite{xia2024fuzz4all}, in which the Test Programmer is instructed to freely modify the given test without guidance.
Following the setup in Section~\ref{ssec:rq1}, three mutated tests were generated for each original test.

The results are presented in Table~\ref{tab:mutation_ablation}.
We observed higher coverage for tests generated by free mutation in four core \textit{fastjson} classes.
This is because more tests generated by agent-written mutation rules fail to compile.
As a result, free mutation produces more valid unit tests, resulting in higher coverage.
Though provided with instructions and examples along the mutation rules, the LLM may not use them correctly in the generated code, resulting in fewer compiled tests.

\begin{figure}[t]
  \centering
  \includegraphics[width=0.85\linewidth]{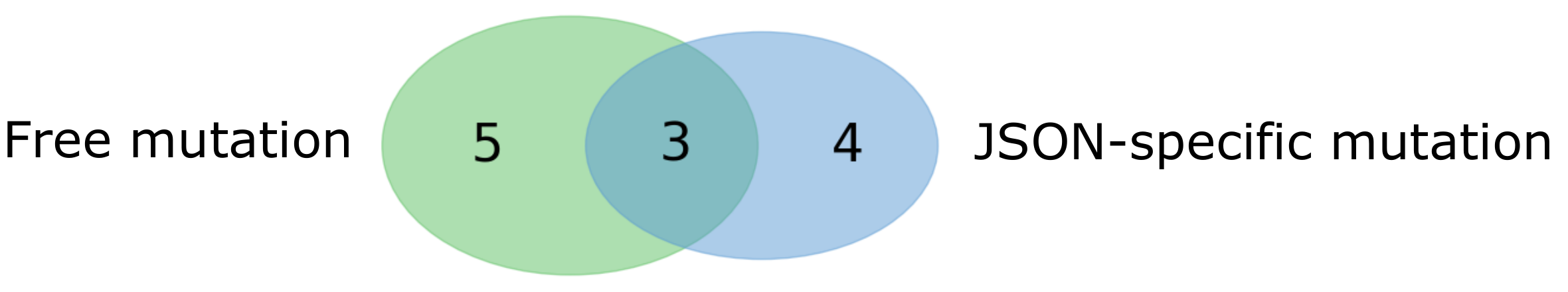}
  \caption{Detected bugs of applying different mutation rules}
  \label{fig:bug_mutation}
\end{figure}

Prompting with free mutation and agent-written mutation rules can both identify bugs, as seen in Fig.~\ref{fig:bug_mutation}.
We observed that both mutation rules (either free or agent-written) are capable of identifying unique bugs, with three bugs overlapping in between.
Additionally, five unique bugs were identified using free mutation, while four unique ones were identified using agent-written mutation.
We found that LLM tends to make more conservative changes to the original tests with free mutation, resulting in the detection of certain unique bugs.
\textbf{However, they were mostly known bugs as confirmed by the developers.}
In contrast, the LLM produced larger changes with agent-written mutation, thereby discovering additional distinct bugs.
This suggests that incorporating such rules into the prompts can help uncover distinct and new bugs that the base setting alone might miss.
In conclusion, applying agent-written mutation rules helps diversify the generated test suite.

\begin{table}[t]
\centering
\caption{Coverage scores of applying different mutation rules}
\label{tab:mutation_ablation}
\begin{tabular}{c c c}
\toprule
        & \multicolumn{2}{c}{Instruction Coverage (\%)} \\
                    \cmidrule(lr){2-3}
Class & Free Mutation & Agent-written Mutation \\ \midrule
\texttt{JSON}       & \textbf{36.14} & 32.76 \\
\texttt{JSONPath}   & \textbf{28.17} & 20.46 \\
\texttt{JSONArray}  & 26.35 & \textbf{29.64} \\
\texttt{JSONObject} & 37.53 & \textbf{41.33} \\
\texttt{JSONReader}     & \textbf{47.10} & 44.86 \\
\texttt{JSONWriter}     & \textbf{43.74} & 41.88 \\
\bottomrule
\end{tabular}
\end{table}

\subsubsection{Effectiveness of Summarizing Test Intention}
\label{sssec:summarization}

To assess the impact of integrating test intention summarization into the Test Programmer's context, we instructed {\tool} to generate three new tests for each bug-triggering test in our dataset, this time omitting the test summarization step.
To control uncertainty, we constrained the Test Programmer to apply only the free mutation rule.
We then compared the resulting test suite with that obtained in Section~\ref{sssec:mutation-rules}.

\begin{table}[h]
\centering
\caption{Test execution result of test summarization (\%)}
\label{tab:summarization_ablation}
\begin{tabular}{c c c}
\toprule
Execution Result & w/ summarization & w/o summarization  \\ 
\midrule
 {Pass} & $56.3$ & $54.2\downarrow$  \\
 {Failure/Exception} & $29.8$ & $20.1\downarrow$ \\
 {Compile Error} & $13.9$ & $25.7\uparrow$ \\
\bottomrule
\end{tabular}
\end{table}

Table~\ref{tab:summarization_ablation} shows that tests generated with intention summarization achieve a higher pass rate (56.3\% vs. 54.2\%) compared to those generated without summarization.
Notably, removing the summarization step leads to a significant increase in compilation errors, rising from 13.9\% to 25.7\%.
Our in-depth investigation revealed that these compilation failures were primarily caused by missing bean class definitions (70\%), missing import statements (20\%), and hallucinated method names (10\%).
Although the failure rate of executable tests is slightly higher with summarized intentions (29.8\% vs. 20.1\%), both passing and failing tests contribute to code coverage and bug detection.
Overall, incorporating intention summarization produces test suites that are more practically useful.

\subsubsection{Performance Variation of Different LLMs}
\label{sssec:effects_llm}

To evaluate the effectiveness of Test Programmer with different LLMs, we experimented with open-source LLMs Llama3 and DeepSeek, and the commercial LLM GPT-3.5.
All models were configured with a \textit{top-p} of 0.95 and a \textit{temperature} of 0.8.
We also prompted them to perform free mutations.
We asked each LLM to generate three tests for each of the original bug-triggering tests and compare their resulting test suite.

\begin{table}[t]
\centering
\caption{Coverage scores of different LLMs for the Test Programmer}
\label{tab:llm_ablation}
\begin{tabular}{c c c c}
\toprule
        & \multicolumn{3}{c}{Instruction Coverage (\%)} \\
                    \cmidrule(lr){2-4}
    {Class}    & {Llama3} & {DeepSeek} & {GPT-3.5}       \\
    \midrule
    \texttt{JSON}       & 28.19  & \underline{36.12} & \textbf{36.14}     \\
    \texttt{JSONPath}   & \underline{19.67}     & 8.22      & \textbf{28.17} \\
    \texttt{JSONArray}  & 22.39  & \underline{22.81}     & \textbf{26.35} \\
    \texttt{JSONObject} & 30.23  & \underline{34.95}     & \textbf{37.53} \\
    \texttt{JSONReader}     & 40.15  & \underline{43.90}     & \textbf{47.10} \\
    \texttt{JSONWriter}     & 36.24  & \underline{41.88}     & \textbf{43.74} \\
\bottomrule
\end{tabular}
\end{table}

The coverage scores of the generated tests are shown in Table~\ref{tab:llm_ablation}.
Among the three models, Llama3 achieved the lowest coverage score.
The reason is that it suffers from a higher compile error rate of $25.71\%$, compared to $13.82\%$ for GPT-3.5 and $12.73\%$ for DeepSeek.
GPT-3.5 achieved the highest coverage, reflecting its superior capabilities to generate accurate and functional code across diverse scenarios.
DeepSeek follows closely behind GPT-3.5, showing a competitive performance in generating code with fewer compile errors.

\begin{figure}[t]
  \centering
  \includegraphics[width=0.85\linewidth]{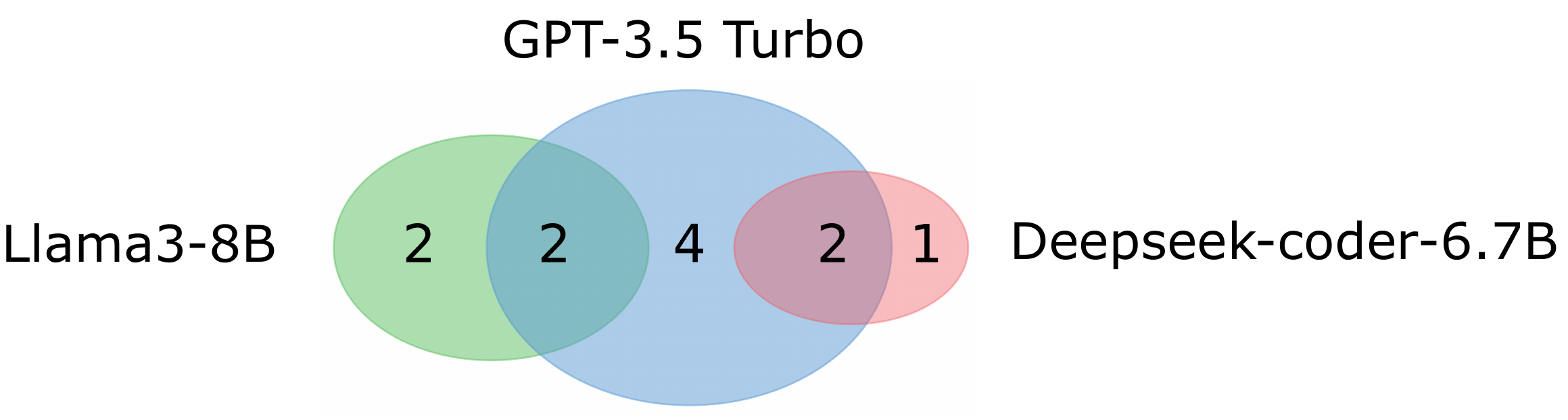}
  \caption{Detected bugs of different LLMs for the Test Programmer}
  \label{fig:bug_llm_venn}
\end{figure}

Fig.~\ref{fig:bug_llm_venn} shows the overlaps between the identified bugs triggered by different LLMs.
Overall, GPT-3.5 identified the most bugs among the models, while both Llama3 and DeepSeek detected unique bugs that were not triggered by GPT-3.5.
While LLMs vary in overall performance, differences in their training data and methodologies enable them to identify distinct, non-overlapping functional defects.
This suggests that, in practice, combining outputs of different models can enhance bug detection and lead to more comprehensive testing.

\begin{center}
\noindent\fbox{\parbox{.9\linewidth}{
\textbf{Answer to RQ2.}
Agent-written mutation rules help uncover unique bugs that the base prompt may miss.
Involving test summarization reduces the compile error rate.
Additionally, the Test Programmer can be adapted to use different LLMs to detect more diverse bugs.
}}
\end{center}

\subsection{RQ3: Detected Bugs}
\label{ssec:rq3}

We allocated the OpenAI API budget to \$25 and ran {\tool} to extensively test the \textit{fastjson} library, consuming approximately 40 CPU hours.
Within this budget, {\tool} revealed 59 bugs in \textit{fastjson} that relate to various features, such as annotations, \texttt{JSONPath} and (de)serialization.
% These bugs were reported to the project's issue tracking system.
A complete list of our reported bugs can be found in \textit{fastjson}'s issue tracking system\footnote{\url{https://www.github.com/alibaba/fastjson2/issues?q=is\%3Aissue\%20state\%3Aclosed\%20author\%3ACooperzzy}}.
% \footnote{The issues were submitted via a personal account and are currently omitted for anonymity.}.
Among the reported bugs, 47 were confirmed by \textit{fastjson}'s developers, and 28 had already been fixed in the latest release.
Here, we present three representative examples of the discovered bugs to demonstrate the practical usefulness of \tool.

% bug 2: WriteNonStringValueAsString with Boolean
\begin{lstlisting}[style=java, caption=A serialization bug with \texttt{Boolean} wrapper class, label=list:boolean_string_bug]
Bean bean = new Bean(Boolean.FALSE);
String json = JSON.toJSONString(bean,
          Feature.WriteNonStringValueAsString);
assertEquals("{\"b\":\"false\"}", json);

public static class Bean { private Boolean b; }
\end{lstlisting}

Listing~\ref{list:boolean_string_bug} illustrates a serialization bug involving the \texttt{Boolean} wrapper class, as previously shown in Fig.~\ref{fig:fuzz_overview}.
The \texttt{JSON.toJSONString} method is invoked to serialize an object, where non-string values should be written as strings (i.e., enclosed in quotation marks).
However, \textit{fastjson} produces the incorrect output \texttt{"\{\textbackslash"b\textbackslash":false\}"} at line 4, omitting the quotation marks around the \texttt{Boolean} value and thereby causing an assertion failure.
This bug was confirmed and discussed in \href{https://github.com/alibaba/fastjson2/issues/2560}{issue 2560}.
Listing~\ref{list:boolean_string_bug} is generated from the test in \href{https://github.com/alibaba/fastjson2/blob/main/core/src/test/java/com/alibaba/fastjson2/issues_1800/Issue1874.java}{issue 1874}.
The Test Programmer, guided by the agent-generated mutation rule, replaces the original configuration option, successfully revealing a serialization bug.

% bug 4: JSONPath.eval with String and Object
\begin{lstlisting}[style=java, caption=A bug in the \texttt{JSONPath} class, label=list:jsonpath_eval]
JSONObject obj = JSONObject.of("data",
                               JSONArray.of(1));
String str = JSON.toJSONString(obj);
assertEquals(JSONPath.eval(str, "$.data[0][0]"), 
             JSONPath.eval(obj, "$.data[0][0]"));
\end{lstlisting}

Listing~\ref{list:jsonpath_eval} demonstrates an example of an inconsistency bug in \texttt{JSONPath}.
The \texttt{.eval()} method produces different results when processing \texttt{JSONObject} and \texttt{String}, even though they contain the same JSON data.
This case leads to a failed assertion at line 4.
The bug was confirmed in \href{https://github.com/alibaba/fastjson2/issues/2584}{issue 2584}.
The original test behind Listing~\ref{list:jsonpath_eval} can be found in \href{https://github.com/alibaba/fastjson2/blob/main/core/src/test/java/com/alibaba/fastjson2/issues_1900/Issue1965.java}{issue 1965}.
It only checks for null values of the \texttt{JSONPath}'s evaluation results.
With this test case, \tool's Test Programmer made a valid modification that directly compares the evaluation results on both the original \texttt{JSONObject} and the serialized string, which successfully triggers an inconsistency bug.

% bug 1: Deserialization with BigDecimal
\begin{lstlisting}[style=java, caption=A deserialization bug with \texttt{BigDecimal} class, label=list:BigDecimalBug]
BigDecimal decimal = BigDecimal.valueOf(
            Long.MAX_VALUE).add(BigDecimal.ONE);
String str = JSON.toJSONString(decimal);
assertEquals("9223372036854775808", str);
BigDecimal dec1 = (BigDecimal)JSON.parseObject(
                         str, BigDecimal.class);
assertEquals(decimal.stripTrailingZeros(), dec1);
\end{lstlisting}

Listing~\ref{list:BigDecimalBug} shows a deserialization bug with the JDK class \texttt{BigDecimal}.
When attempting to parse a JSON string into a \texttt{BigDecimal}, the \texttt{parseObject} method incorrectly interprets it as negative, returning $-9223372036854775808$ and triggering an assertion failure at line 7.
This bug was confirmed in \href{https://github.com/alibaba/fastjson2/issues/2582}{issue 2582}.
The original test case behind this bug appears in  \href{https://github.com/alibaba/fastjson2/blob/main/core/src/test/java/com/alibaba/fastjson2/issues_1000/Issue1204.java}{issue 1204}.
In this case, the Test Generator uses the \texttt{parseObject} method instead of the original \texttt{parse}, thereby uncovering a bug related to numeric overflow.

By comparing the differences between the original and newly generated tests, we conclude that {\tool} effectively enhances the diversity of historical bug-triggering test cases, thereby facilitating the discovery of new bugs.

\begin{center}
\noindent\fbox{\parbox{.9\linewidth}{
\textbf{Answer to RQ3.}
{\tool} effectively facilitates the discovery of 59 reported bugs in \textit{fastjson}, of which 47 were confirmed and 28 have already been fixed.
}}
\end{center}

\subsection{RQ4: Failures Analysis}
\label{ssec:rq4}

\subsubsection{Failures of Generated Tests}

In Section~\ref{ssec:rq1}, among all of the \tool-generated tests that compiled, 67.0\% passed, 25.6\% failed due to assertion mismatches, and 7.4\% crashed due to other exceptions or errors.
We inspected the failing and crashing tests to identify \textbf{if they were bad tests} and investigate their root causes.

For tests that fail at assertions, the most common cause is the assertion of contradictory expected values.
Typical scenarios of such failing test assertions include:
1) expecting field names that contradict the bean fields;
2) wrongly asserting a null or non-null object; or
3) asserting incorrect data types.
Another common issue is incorrect expected values of the serialized string, such as the one shown in Fig.~\ref{fig:badtest_example}.

The top three exceptions thrown are: \texttt{JSONException}, \texttt{\small NullPointerException}, and \texttt{ClassCastException}.
The primary cause is that parsing an incorrect JSON string, either syntactically incorrect or mismatched with the bean definition, directly results in a library-defined \texttt{JSONException}.
Other reasons include calling parsing/serializing functions with incorrect parameters, casting data to incompatible types, and invoking methods on a null object that was retrieved from a non-existent key or index.

\subsubsection{Bad Test Classification by Test Validator}
\label{sec:test_classification}

Our investigation above confirmed that the test cases generated by the Test Programmer do include bad tests, highlighting the need for the Test Validator agent in the overall framework.
Therefore, this subsection evaluates the classification accuracy of the Test Validator in isolation.
We collected 21 bad tests and 22 good tests.
Among them, 10 bad tests triggered exceptions ($E_\textit{bad}$), 11 bad tests failed at assertions ($F_\mathit{bad}$), 10 good tests triggered exceptions ($E_\textit{good}$), and 12 good tests failed at assertions ($F_\mathit{good}$).
Instances of $E_\textit{good}$ and $F_\textit{good}$ were collected from test cases that have identified bugs, and we sampled a similar number of $E_\textit{bad}$ and $F_\textit{bad}$ from the generated tests to ensure balanced sample sizes.
We then ran Test Validator using the prompt in Fig.~\ref{fig:test_validator} and recorded the results, as shown in Table~\ref{tab:classify_result}.

\begin{table}[t]
\centering
\caption{Bad Test Classification Results}
\label{tab:classify_result}
\begin{tabular}{l c c c c c}
    \toprule
    & $E_\textit{bad}$ & ${E}_\textit{good}$ & ${F}_\textit{bad}$ & $F_\textit{good}$ & Accuracy (avg.) \\
    \midrule
    FS  & 7/10 & 4/10 & 10/11 & 10/12 & 72.1\% \\
    FS-CoT & 5/10 & 5/10 & 10/11 & 10/12 & 69.8\% \\
    \bottomrule
\end{tabular}
\end{table}

For bad test samples that fail due to assertion mismatches (${F}_{bad}$ and ${F}_{good}$), the Test Validator demonstrates a stronger ability to identify them.
However, for cases that throw exceptions ($E_{bad}$ and $E_{good}$), its performance appears closer to random guessing.
This can be attributed to LLM's ability to reason about code behavior, while its difficulty lies in predicting library-defined exceptions (e.g., \texttt{JSONException} in \textit{fastjson}).
Such exceptions can be triggered by both invalid test logics (bad tests) and internal library issues (good tests), making accurate classification more challenging.
Interestingly, the few-shot with chain-of-thought (FS-CoT) approach, which includes reasoning steps in the few-shot examples, performs similarly to the basic few-shot (FS) method.

Among the 14 misclassified instances by both FS and FS-CoT, 11 are shared between the two approaches.
Many of these misclassifications are related to specific behaviors of \textit{fastjson}, which are inherently difficult to judge even for experienced users~\cite{issue3153_judge}.
As such, even with explicit reasoning steps, accurately assessing new cases remains challenging without prior domain-specific knowledge.
Nevertheless, the experimental results indicate that the Test Validator is still able to identify a meaningful portion of bad tests, contributing to the quality improvement of \tool's generated test suites.

\begin{center}
\noindent\fbox{\parbox{.9\linewidth}{
\textbf{Answer to RQ4.}
Most bad tests come from assertion mismatches and library-defined exceptions.
While Test Validator effectively identifies bad tests that fail at assertions, it struggles to accurately classify those that trigger library-defined exceptions.
}}
\end{center}

\section{Discussions}

\subsection{Lessons Learned}

\noindent$\bullet$ \textbf{Existing general-purpose LLM-based ATG approaches are better at generating regression test suites than bug-triggering tests.}
In our experiment, most test cases produced by \textsc{ChatTester} and \textsc{ChatUniTest} are short and trivial.
They generally consist of straightforward method invocations and object initializations without complex data manipulation operations.
Such tests can easily achieve a promising coverage score and thus are suitable for regression purposes.
However, these tests rarely uncover edge cases or unusual input, limiting their practical use for bug finding.

\noindent$\bullet$ \textbf{Leveraging different LLMs can enhance the diversity of generated tests, thereby improving the bug detection capability.}
By using different LLMs, such as open-source (Llama3), coding-specific (DeepSeek-coder), and commercial (GPT-3.5) LLMs, {\tool} uncovers a broader range of bugs.
Differences in pre-training and fine-tuning phases lead to varied generation styles for different models, contributing to diverse generated tests.
Notably, in our experiment, a smaller model (with 6.7B parameters) can also perform well in generating bug-triggering tests for \textit{fastjson}, comparable to larger commercial models like GPT-3.5.

\noindent$\bullet$ \textbf{Explicitly importing expert knowledge enhances LLM-based ATG in specialized domains.}
Throughout our research, we found that providing domain-specific context—such as curated API documentation, known usage patterns, and historical bug-triggering tests—can improve the relevance and correctness of the generated test cases.
Rather than relying solely on the general-purpose capabilities of the LLM, encoding expert knowledge into the prompt or system design guided the model to produce more meaningful and valid outputs.
This is particularly important for specialized libraries, where correct usages often depend on complex serialization behaviors that are not easily inferred from natural language alone.
Our results suggest that LLM-based tools for automated testing can benefit greatly from structured expert input, especially in domains with non-trivial semantics or specialized APIs.

\subsection{Limitations and Future Work}

\noindent$\bullet$ \textbf{Uncompiled and bad generated test cases.}
{\tool} may produce test cases that fail to compile, which wastes computational resources and potentially loses otherwise valuable tests.
To address this issue, we apply a series of lightweight post-processing steps to the outputs of \tool's Test Programmer.
Despite these efforts, a significant number of test cases remain uncompiled and are discarded.
One potential solution is to incorporate more fine-grained domain knowledge (e.g., valid examples of Java JSON library usages) into the prompts, which may help reduce such errors~\cite{guan2024large}.
Similarly, test cases labeled as ``bad tests'' by the Test Validator are also discarded.
For these, applying automatic program repair techniques—such as LLM-based self-debugging—offers a promising direction for reusing the discarded tests~\cite{xia2023automated,chen2024teaching}.

\noindent$\bullet$ \textbf{False positives of Test Validator's prediction.}
The generated tests may contain logical errors; therefore, we introduce the Test Validator to identify and filter out such bad test cases.
However, as shown in our experiments, the Test Validator may produce false positives.
To support practical bug detection, we employed a differential oracle~\cite{deng2024large} between the \textit{fastjson2} library and its predecessor, \textit{fastjson}, to isolate true bugs.
However, this approach requires multiple implementations of the same specification, which may not hold in general development scenarios.
As future work, instead of relying on the Test Validator, we plan to explore test migration~\cite{gao2024mut} as a way to facilitate differential testing.
This would involve comparing semantically equivalent tests across various Java JSON libraries.
The challenge here lies in the differences between library APIs—some APIs have direct counterparts in other libraries, while others are specific to a particular library, lacking equivalent functionality elsewhere.
A possible direction is to integrate the Test Programmer agent with a reasoning model to form a joint test oracle, further enhancing the testing process and generalizability across different Java JSON libraries.

\subsection{Threats to Validity}

The main threat to validity lies in the inherent randomness of the LLM agents' outputs.
To evaluate the effectiveness of LLMs, we experimented with both open-source models (Llama3 and DeepSeek) and a commercial model (GPT-3.5) for the Test Programmer agent.
To ensure stable output, we fix generation parameters such as \textit{temperature} and \textit{top-p} sampling, and use specific versions of each agent's underlying LLM.

In \tool's design, we assume that the JSON-P and JSON-B standards apply to the tested Java JSON libraries; however, this assumption may not hold universally across all APIs and libraries.
To mitigate this threat, we use these standards as references for common usage patterns rather than as strict API dependencies.
Moreover, by incorporating the original test case, {\tool} adapts to the specific usage pattern of the target library, allowing it to remain effective even when the library partially deviates from the standard APIs.

\section{Related Work}

\subsection{Traditional ATG Approaches}

ATG approaches \cite{serra2019effectiveness} aim to extensively exercise the software under test by exploring diverse combinations of execution paths, running configurations, test inputs, etc.
Existing studies typically addressed the challenges of mitigating combinatorial explosions during software executions.
Before the emergence of LLM, traditional approaches often employed random-based~\cite{pacheco2007randoop} or search-based~\cite{fraser2011evosuite} strategies to synthesize test programs, or adopted model-based algorithms to accomplish comprehensive functional coverage~\cite{enoiu2016automated}.
A notable example is \textsc{EvoSuite}~\cite{fraser2011evosuite}, which applies an evolutionary algorithm to generate test cases that maximize code coverage.
Although achieving good coverage measurement, these approaches often produce tests that lack readability and maintainability, making them difficult for developers to use in practice \cite{almasi2017industrial,yang2024evaluation}.
More recent research has explored using deep learning techniques to improve ATG \cite{dinella2022toga,watson2020learning}.

Fuzzing \cite{manes2019art} with program libraries often means generating complete programs or fuzzing harnesses that cover various library APIs, which is similar to ATG.
RULF~\cite{jiang2021rulf} generates fuzz targets in Rust libraries by traversing their API dependency graph.
Rahalkar presents a system that generates fuzzing harnesses for library APIs and binary protocol parsers by analyzing unit tests~\cite{rahalkar2023automated}.
Chen et al.~\cite{chen2023hopper} present a general fuzzer for testing libraries without necessitating domain-specific knowledge in creating fuzz drivers.

These techniques have been used for testing JSON libraries.
For example, the JSON fuzzer of \textit{fastjson2} in OSS-Fuzz~\cite{oss-fuzz} targets a specific parsing method, while \textsc{EvoGFuzz}~\cite{eberlein2020evolutionary} generates inputs for eight JSON parsers to uncover defects.
However, they did not specifically target Java JSON libraries, leaving a wide range of common functionalities untested, and {\tool} fills this gap.

\subsection{LLMs for ATG}

The emergence of LLMs has enabled ATG through prompting with natural language.
For instance, FuzzGPT~\cite{deng2024large} and TitanFuzz~\cite{deng2023large} utilize LLMs to generate Python code for testing deep learning libraries.
Fuzz4All~\cite{xia2024fuzz4all} is an LLM-based universal fuzzer for multiple programming language infrastructures.
Yuan et al.~\cite{yuan2024evaluating} conducted an empirical study on ChatGPT’s ATG abilities and introduced the \textsc{ChatTester} tool.
\textsc{TestPilot}~\cite{schafer2023empirical} generates unit tests for JavaScript methods within project APIs. 
\textsc{ChatUniTest}~\cite{chen2024chatunitest} provides an LLM-empowered framework with user-friendly APIs for ATG.
Meta Inc. uses LLMs to automatically improve existing human-written tests and has been deployed at scale in production~\cite{alshahwan2024automated}.
\textsc{UTGen}~\cite{deljouyi2025leveraging} combines search-based testing with LLMs to improve the understandability of the generated unit tests.
While they have shown promise in generating high-coverage tests, challenges remain in ensuring the correctness of the generated tests, as LLM outputs can be non-deterministic and prone to hallucinations.

Unlike previous single-LLM approaches, {\tool} orchestrates three specialized LLM agents to enable end-to-end ATG.
By decomposing the overall task into well-defined subtasks and assigning each to a dedicated agent, {\tool} emulates the workflow of an expert human tester, that is:
1) analyzing historical bug-triggering tests and library specifications to derive meaningful mutation rules;
2) interpreting the intended test behavior and generating new tests guided by the rule; and
3) inspecting the generated tests to assess their validity.

\section{Conclusion}

We present \tool, a multi-agent ATG system specialized for Java JSON libraries.
Leveraging high-quality historical bug-triggering tests and LLM-based agents for mutation rule generation, code summarization, test generation, and validation, {\tool} produces diverse and valid test cases that reveal both crashing and non-crashing bugs in the widely used Java JSON library.
Our experiments on \textit{fastjson} show that {\tool} outperforms two state-of-the-art LLM-based ATG tools in terms of code coverage and detected bugs.
It also demonstrates practical usefulness by reporting 59 bugs, of which 47 were confirmed and 28 have already been fixed.
These results underscore \tool’s potential to advance automatic quality assurance in the JSON domain.

\section*{Acknowledgment}

We would like to thank all anonymous reviewers for their valuable comments on this paper.
This work is supported by the National Natural Science Foundation of China (Grant No. 62372219).

\IEEEtriggeratref{40}
\bibliographystyle{IEEEtran}
\bibliography{reference,web}

\end{document}